\documentclass[a4paper,onecolumn,twoside,fleqn,10pt,runningheads]{article}
\usepackage{amsmath}
\usepackage{authblk}
\usepackage[svgnames]{xcolor}
\definecolor{blue}{RGB}{0,0,255}
\usepackage{caption}
\usepackage[font={color=blue},figurename=Fig.,labelfont={it}]{caption}
\usepackage{bm}
\usepackage{subcaption}
\usepackage[varg]{txfonts}
\usepackage{graphicx}
\usepackage{pdfpages}
\usepackage{wasysym}
\usepackage{rotating}
\usepackage{hyperref}
\usepackage{indentfirst}
\usepackage{float}
\usepackage{lipsum} 
\usepackage{setspace} 
\usepackage{indentfirst}
\usepackage[english]{babel}
\usepackage[switch,columnwise]{lineno}
\usepackage[left=2cm,right=2cm,top=2cm,bottom=2cm]{geometry}

\hypersetup{
    colorlinks=true,                          
    linkcolor=blue, 
    citecolor=blue, 
    urlcolor=blue,
    linktoc=all
}

\newcommand{\blue}[1]{\textcolor{blue}{#1}}

\newcommand{\eg}{\textit{eg }}
\newcommand{\cf}{\textit{cf }}

\newcommand{\Milankovic}{\blue{Milankovic} }

\usepackage{fancyhdr}
\begin{document}
%%%%%%%%% TITLE
\title{On the external forcing of global eruptive activity in the past 300 years}
%%%%%%%%% AUTHORS
\author[1]{J-L. Le Mouël}
\author[2]{D. Gibert}
\author[1]{V. Courtillot}
\author[3,4]{S. Dumont\thanks{ssdumont@fc.ul.pt}}
\author[5]{J. de Bremond Ars}
\author[6]{S. Petrosino}
\author[7]{P. Zuddas}
\author[1]{F. Lopes}
\author[8]{J-B. Boul\'e}
\author[9]{M.C. Neves}
\author[3]{S. Cust\'odio}
\author[3,10]{G. Silveira}
\author[11]{V. Kossobokov}
\author[12]{L. Coen}
\author[13]{M. Geze}

\affil[1]{Universit\'e Paris Cité, Institut de Physique du globe de Paris, CNRS UMR 7154, F-75005 Paris, France}
\affil[2]{LGL-TPE - Laboratoire de Géologie de Lyon - Terre, Planètes, Environnement, Lyon, France}
\affil[3]{Universidade de Lisboa, Faculdade de Ciências, Instituto Dom Luiz (IDL), Lisboa, Portugal}
\affil[4]{Universidade da Beira Interior, Instituto Dom Luiz (IDL), Covilhã, Portugal}
\affil[5]{Universit\'e de Rennes, CNRS, G\'eosciences Rennes - UMR 6118, Rennes, France}
\affil[6]{Instituto Nazionale di Geofisica e Vulcanologia, Sezione di Napoli – Osservatorio Vesuviano, Naples, Italy}
\affil[7]{Sorbonne Universit\'e, CNRS, METIS, F75005, Paris, France}
\affil[8]{CNRS UMR7196, INSERM U1154, Museum National d'Histoire Naturelle, Paris, F-75005,  France}
\affil[9]{Universidade do Algarve, FCT, Campus de Gambelas, 8005-139, Faro, Portugal }
\affil[10]{Instituto Superior de Engenharia de Lisboa, Rua Conselheiro Emídio Navarro 1, 1959-007 Lisboa, Portugal}
\affil[11]{Institute of Earthquake Prediction Theory \& Mathematical Geophysics, Russian Academy of Sciences (IEPT RAS), Moscow, Russian Federation}
\affil[12]{Mus\'eum National d’Histoire Naturelle, /CNRS laboratoire PHYMA, UMR7221, F-75005, France}
\affil[13]{Mus\'eum National d’Histoire Naturelle, Sorbonne Université, CEMIM, Paris, F-75005, France}

\date{\today}
\maketitle
\ \\
\newpage
\ \\

%%%%%%%%% ABSTRACT
\abstract {The decryption of the temporal sequence of volcanic eruptions is a key step in better anticipating future events. Volcanic activity is the result of a complex interaction between internal and external processes, with time scales spanning multiple orders of magnitude. We review periodicities that have been detected or correlated with volcanic eruptions/phenomena and interpreted as resulting from external forces. Taking a global perspective and longer time scales than a few years, we approach this interaction by analyzing three time series using singular spectral analysis: the global number of volcanic eruptions (NVE) between 1700 and 2022, the number of sunspots (ISSN), a proxy for solar activity, the polar motion (PM) and length of day (lod), two proxies for gravitational force. Several pseudo-periodicities are common to NVE and ISSN, in addition to the 11-year Schwabe cycle that has been reported in previous work, but NVE shares even more periodicities with PM. These quasi-periodic components range from ~5 to ~130 years. We interpret our analytical results in light of the Laplace’s paradigm and propose that, similarly to the movement of Earth's rotation axis, global eruptive activity is modulated by commensurable orbital moments of the Jovian planets, whose influence is also detected in solar activity.}

\section{Introduction\label{sec:01}}
	Volcanic eruptions represent the ultimate expression of a long series of physico-chemical processes initiated in the mantle with the generation of magma. Isotopic studies, mainly based on U-series, have shown that the formation, accumulation, and evolution of magma are processes that occur over long time scales, typically ranging from a few hundred years to a few millennia. On the other hand, magma transport, ascent to the surface, and other processes leading to eruption, as identified by geophysical and petro-geochemical studies, are much faster, occurring over days to decades (see \cite{Cashman2013}).  Closely linked to plate tectonics and mantle dynamics, volcanic activity can also be seen as a response to changes in stress fields resulting from regional or global processes, such as large tectonic earthquakes or tidal effects from the Earth and oceans. Thus, volcanic eruptions reflect the internal dynamics of our planet, often acting as a "regulator" of accumulated stresses in the lithosphere over a wide range of time scales, from decades to centuries and beyond, as illustrated by rifting episodes along divergent plate margins (\eg \cite{Bjornsson1985}) or the effect of variations in ice cap thickness on magma plumbing systems (\eg \cite{Watt2013,Lucas2022}). Deciphering the temporal pattern of volcanic activity is therefore crucial for assessing triggers of volcanic eruptions. This would not only help understand the dynamics of their internal plumbing system, but also unravel the interaction with potential external processes that may contribute to the destabilization of volcanoes on variable time scales  (\eg \cite{Dzurisin1980,Mcnutt1987,Kutterolf2013,Watt2013,Canon2014,Coussen2016,Watt2019,Seropian2021,Bilham2022,Dumont2022,Lucas2022}). Furthermore, the recognition of specific temporal patterns in eruptive activity is a crucial objective in volcanology, as it would allow for anticipation of future events and thus better mitigation of volcanic hazards.

	The coverage of volcanic eruptions has become increasingly systematic and comprehensive over the past two centuries  (\eg  \cite{Simkin1993,Gusev2008}). However, for most individual volcanic systems, only a few dozen events are recorded, making any statistical study of local structures at this timescale particularly challenging. Nevertheless, when considered collectively for regional or global analysis, volcanic eruptions can reveal specific spatial and temporal patterns that can ultimately be correlated and linked to global processes. This approach has already contributed to demonstrating that the spatial and temporal distribution of eruptive events can be related to sea level variations (\eg \cite{Mason2004,Kutterolf2013}). Recently, \blue{Dumont et al.} (\cite{Dumont2022}) showed that the major decadal periodicities present in both the global mean sea level and the monthly number of eruptions are also present in the movement of Earth's rotation axis. This has led the authors to hypothesize that there could be a causative link between pole motion (i.e., movement of Earth's rotation axis), a global process, and local events, i.e., volcanic eruptions occurring worldwide. The redistribution of fluid masses in the crust would act as the coupling link between these two processes; it would enable the transfer of stresses, a link already suggested at seasonal scale (\eg \cite{Mcnutt1987,Mason2004}), and over several thousand years  (\eg \cite{Kutterolf2013,Kutterolf2019,Satow2021}).
   
	Some studies (some dating back over a century) have highlighted an apparent correlation between global, regional, and local volcanic activity and solar cycles (\eg \cite{Lyons1899,Jensen1902,Stothers1989}). Similar observations have been made for large earthquakes as well (\eg \cite{Jensen1902,Mazzarella1988,Mazzarella1989,Marchitelli2020}). Furthermore, it appears that the occurrence of some of these major events, whether seismic or volcanic (such as the eruption of Pinatubo in 1991), is associated with specific planetary alignments (\eg \cite{Greco2021,Safronov2022}).However, most of these correlations/associations between local geological phenomena and global processes such as solar activity or planetary alignment are widely debated (\eg \cite{Akhoondzadeh2022}), primarily due to a major challenge: the lack of a credible physical mechanism (\eg \cite{Jensen1902,Stothers1989,Marchitelli2020}). How do these interactions work and why do only certain geological structures seem to be sensitive to such external influences?

	In the following, we revisit these correlations between volcanic eruptions and global processes from a global perspective, in light of recent studies (\cite{Courtillot2021,LeMouel2021a,Lopes2021a,Bank2022,Scafetta2022,LeMouel2023,Scafetta2023}). We address the interaction between local eruptive phenomena occurring worldwide on planet Earth, i.e., the global number of volcanic eruptions (\cite{venzke2013}), and global processes using a two-step approach. First, we analyze two time series: the global number of volcanic eruptions (NVE) and solar activity recorded by the international sunspot number (ISSN). Then, we perform a similar analysis/comparison with Earth's pole motion (PM), considering that a link between the coordinates of the rotation pole, sea level variations, and global volcanic eruptions has already been introduced in \blue{Dumont et al.} (\cite{Dumont2022}). These analyses involve extracting, identifying, and comparing the common quasi-periodic components present in the different time series since 1700. This two-step procedure first allows us to assess whether volcanic eruptions on Earth are consistent with Laplace's global theory, the main principles of which are summarized in Section 3; and second, to shed light on a possible common external force that could influence processes occurring in the Sun and in the Earth. In the latter case, the common external force could participate in the processes leading to volcanic eruptions and act as a trigger through a global mechanism of mass redistribution and stress. It is important to note that we do not hypothesize a solar force on volcanic activity, unlike a number of previous studies  (\eg. \cite{Mazzarella1988,Mazzarella1989,Stothers1989,Strestik2003,Herdiwijaya2014,Vasilieva2020}).
	
\section{Periodicities detected in volcanic activity and other geophysical phenomena \label{sec:02}}
	Periodic behaviors have been detected in short-term studies on volcanoes, using either physical parameters or recorded visual observations during periods of unrest, eruptions, and/or relative quiescence (see \cite{Dumont2022} and therein references). Some of these periodicities correspond to those of terrestrial and oceanic tides and have been observed in various records, such as changes in eruptive behavior and explosion rates  (\eg \cite{Sottili2012}), seismic tremor (\eg  \cite{Custodio2003,Dumont2020,Dumont2021,Petrosino2022}), seismicity (\eg \cite{Petrosino2018}), deformation (\eg \cite{deLauro2018}), gas fluxes (\eg \cite{Dinger2018,Dumont2021}), flow of lava at surface -- energy radiated and volume emitted (\eg \cite{Dumont2020,Dumont2021}). These observations show that magmatic and hydrothermal fluids, as well as the volcanic structure, primarily respond to tidal forces with oscillatory behavior ranging from hours to months. However, so far, the available observations have been too scattered to deduce the main conditions that make a volcano more sensitive or capable of responding to lunisolar tides. Nevertheless, a critical state, open or fluid-rich systems represent the main characteristics of volcanoes for which tidal influence has been reported.

	On longer timescales, the triggering of volcanic eruptions, typically considered through the eruptive history of individual volcanoes or analyzed in the context of global or regional catalogs, may also be influenced by lunar cycles, particularly the fortnightly cycle (\eg \cite{Machado1960,Mauk1973,Hamilton1973,Dzurisin1980}). The signature of gravitational forces induced by the relative motion of the Moon and the Sun at these timescales is also present in global parameters recorded on Earth, such as pole motion (\eg \cite{Chao2014,LeMouel2019b}),  and local parameters that respond to global processes on Earth, such as the geomagnetic field (\cite{Courtillot1988,LeMouel2023}), atmospheric pressure and sea level (\cite{Wunsch1972,Chelton1986,Lambeck2005,Courtillot2022a,Lopes2022b}), or tectonic plate motion (\cite{Zaccagnino2020,Zaccagnino2022}). 

	Seasonal variations correspond to oscillations that occur over a year or half a year. They are associated with the Earth's revolution around the Sun. Seasonal variations have been observed in eruptive sequences, both at the level of individual volcanoes and regionally (\cite{Mcnutt1987,Mason2004}). Their correlation with sea level variations has led to the interpretation that some volcanic eruptions are the result of redistribution of oceanic mass. The influence of the Sun (through variations in sunspot numbers) on eruptive activity has also been suggested over longer timescales. Some eruptive events or episodes seem to occur during periods of solar minimum or maximum activity, with cycles mainly of $\sim$11 yr and $\sim$22 yr, but also $\sim$80 yr and more (\eg \cite{Jensen1902,Lyons1899,Machado1960,Schneider1975,Stothers1989,Strestik2003,Khain2007,Qu2011,Casati2014,Herdiwijaya2014,Ma2018, Vasilieva2020,Dumont2022}). Similarly to short-term timescales and tidal periodicities, these multi-annual to centennial variations are present in other geophysical parameters. Most of them record both tidal and solar periodicities, as illustrated by the magnetic field (\eg \cite{Bartels1932,LeMouel1984,Courtillot1988,Jault1991,LeMouel2019a,LeMouel2023}), polar motion (\eg \cite{Lopes2017,Lopes2021a,Lopes2022a}); sea level observed from local and global data (\eg \cite{Chambers2012,Wahl2015,LeMouel2021a,Courtillot2022a}), global surface air temperatures (\eg \cite{Courtillot2013,LeMouel2020b,Scafetta2021,Scafetta2023}) and more generally, climate (\eg \cite{Wood1974,Schlesinger1994,Scafetta2010,Courtillot2013,Scafetta2016,LeMouel2019c,LeMouel2021b}). Moreover, these links also seem to exist 1) on much longer timescales of millennia, governed by astronomical configurations and impacting climate and sedimentation (\eg \cite{Milankovic1920,Kutterolf2013,Coussen2016,Kutterolf2019,Lopes2021b,Satow2021}); and 2) also for earthquakes, as suggested by numerous studies (\eg \cite{Tamrazyan1968,Mazzarella1988,Marchitelli2020}).

	In this section, we sought to highlight the fact that among the many geophysical observables associated with the various processes occurring on and within the Earth (see also \cite{Lopes2021a}), volcanic phenomena are no exception. Although still poorly understood, volcanic phenomena are influenced by periodicities across a wide range of time scales, as is the case for many other geophysical observables measured on Earth, as suggested by various methodologies - computational approach (\eg \cite{LeMouel1984,Jault1991}), statistical (\eg \cite{Mason2004}), spectral analysis (\eg \cite{Scafetta2010}), or decomposition approach (\eg \cite{Lopes2017,LeMouel2019b}). This first-order analogy between the periodic components detected in volcano-related parameters and other geophysical parameters only pertains to the observations of these periodicities and not the interpretations drawn from them. In fact, although lunar-solar gravitational forces and solar activity have both been considered as external forcings acting on the dynamics of our planet and particularly its fluid envelopes, the way they interact is still debated, as we will see in the next section.
	
\section{On Laplace’s theory \label{sec:03}}	
\begin{figure}[H]
       \begin{center}
         \includegraphics[width=0.5\columnwidth]{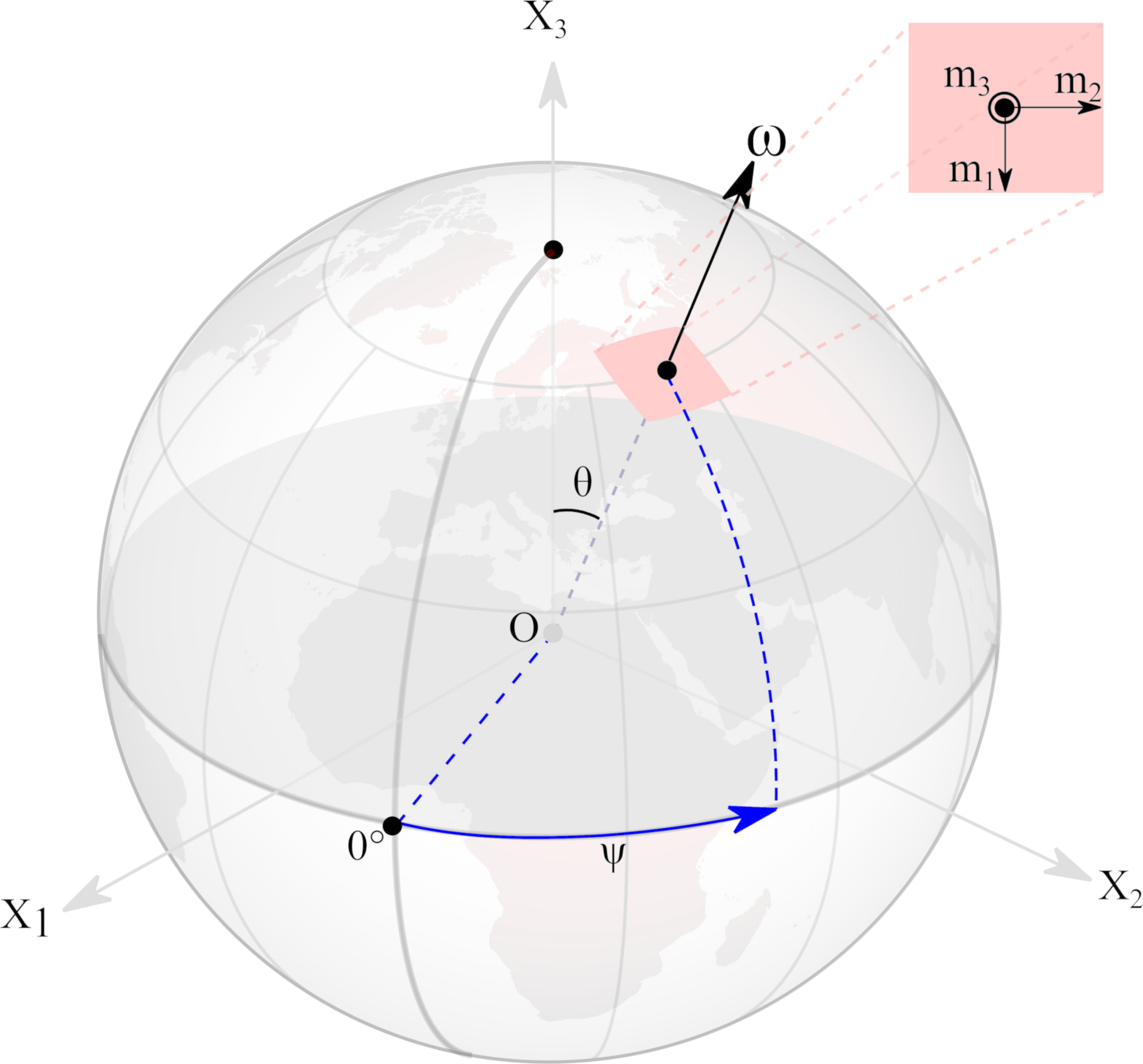}	
         \caption{Geophysical reference system  where the different parameters of the system of Liouville-Euler equations are presented (see system \ref{eq:01} in main text).}
       \end{center}
         \label{Fig:01}
\end{figure}

	In his treatise on celestial mechanics, \blue{Laplace} (\cite{Laplace1799}) asserted that all masses on the surface and interior of the Earth are influenced by orbital moments, also known as angular moments, generated not only by the Moon and the Sun, but also by other celestial bodies present in the solar system. The same applies to the complete motion of the Earth's rotation axis. This theory has been revisited and validated in light of recent analyses based on time series data spanning approximately 300 years of the Earth's rotation axis motion (\cf \cite{Lopes2021a,Lopes2022a,Lopes2023}). This recent theory contradicts the currently most accepted theory (\eg \cite{Lambeck2005}), which attributes a key role to atmospheric dynamics (\cite{Brzezinski2002}). In this latter theory, seasonal oscillations present in many datasets, including polar motion (\eg \cite{Lambeck2005,LeMouel2021a,Courtillot2022b,Lopes2022b}),  have generally been considered in terms of angular momentum and interpreted as resulting from mass redistribution of the atmosphere on the rotating Earth. This current theory has not yet been quantitatively validated (see \cite{Lambeck2005}, chapter 7 introduction). Both of these theories can be considered as a transfer of energy from an undefined energy source to the fluid and solid envelopes of the Earth, as described by the Liouville-Euler equations system (\cf system \ref{eq:01}, \eg \cite{Lambeck2005} chapter 3). The direct implication of this description is that any changes in the motions of the Earth's rotation axis and the length of day (lod) inevitably lead to a reorganization of mass on the surface and interior of the Earth, and thus a redistribution of stresses on the Earth's surface.
\begin{subequations}
\begin{align}
	j (\dot{\textbf{m}}/\sigma_r) + \textbf{m} &= \boldsymbol \psi \\
	\dot{m}_{3} &= \psi _3
\end{align}	
\label{eq:01}
\end{subequations}
	In the system of Liouville-Euler equations (\ref{eq:01}), $j$ represents the complex number such $j^{2}=-1$, $\sigma _r = \dfrac{C-A}{A}\Omega$ is the Euler's free wobble frequency, $A$ ($\sim 8.010 * 10^{37} kg.m^{2}$) and $C$ ($\sim 8.036 * 10^{37} kg.m^{2}$) are respectively the equatorial and axial Earth's momenta, $\Omega$ ($\sim 7.292115*10^{-5} rad.s^{-1}$) is the instantaneous rotation velocity. The $\textbf{m} = m_1+j*m_2$ and $m_3$ parameters, are dimensionless numbers defining the motion of the rotation pole (Figure \ref{Fig:01}); the functions $\boldsymbol \psi = \psi _1 + j*\psi _2$ and $\psi _3$ corresponding to excitation functions (\eg \cite{Lambeck2005}, chapter 3). The reorganization of the masses and stresses on the Earth’s surface intervene through the functions $\boldsymbol \psi$ as illustrated by the description of the isostasy theory (\eg \cite{Nakiboglu1980}).

	This is about identifying common periodic components present in the various envelopes of the Earth that enable the evaluation of these two theories. These periodicities can be divided into two main categories. First, periodicities shorter than 1 year (except for lunar nutation which has a period of 18.6 years). These periods have been attributed to tides induced by luni-solar gravitational forces. Second, periodicities longer than 1 year. The most frequently detected periods in this group are approximately $\sim$11, $\sim$22, $\sim$30, $\sim$90 and $\sim$160 years, many of which are common to solar physics (\eg \cite{Milankovic1920,Gleissberg1939,Gleissberg1944,Jose1965,Coles1980,Courtillot1988,Charvatova1991,Frick1997,LeMouel2017,Usoskin2017,LeMouel2020a,Courtillot2021}). In particular, they have been associated with the Schwabe, Hale, an unnamed 30-year cycle linked to Saturn, Gleissberg, and José solar cycles, respectively (\cf \cite{Schwabe1844,Hale1919,Gleissberg1939,Jose1965}).
	
	\blue{Mörth and Schlamminger} (\cite{Morth1979}) show that these different solar cycles can also be recognized in the orbital motions (specifically, the orbital moments) of pairs and groups of pairs of the four Jovian planets; results that have been confirmed by several recent studies (\eg \cite{Bank2022,Lopes2023}). 
	
	Let us explore this idea by first considering the Sun. Located at the center of the solar system, it directly exerts on all planets of the solar system the strongest gravitational attraction, yet it does not carry any orbital momentum. According to the law of action-reaction in a Galilean referential, we can convert dimensionally the value of Earth’s force of attraction exerted by the Sun ($F = 3.57*10^{22} kg.m.s^{-2}$) into an angular momentum ($F$*Sun-Earth distance*Earth’s revolution period), which corresponds to the angular momentum exerted by the Sun on the Earth, and which is of $1.68*10^{41} kg.m^{2}.s^{-1}$. This value is smaller by 1-3 orders of magnitude than that of the Jovian planets which are of $1.93*10^{43} kg.m^{2}.s^{-1}$ for Jupiter (revolution: $\sim$11 years), $7.82*10^{42} kg.m^{2}.s^{-1}$ for Saturn (revolution: $\sim$30 years), $1.69*10^{42} kg.m^{2}.s^{-1}$ for Uranus (revolution $\sim$90 years) and $2.50*10^{42} kg.m^{2}.s^{-1}$ for Neptune (revolution: $\sim$165 years). \blue{Mörth and Schlamminger} (\cite{Morth1979}) studied this physical quantity through the "commensurability" of different pairs and groupings of planet pairs that make up our solar system. This quantity, which should be rational, represents the ratio of the average orbital period of one or more pairs of planets. It highlights their ability to transmit energy through angular momentum and is associated with periodicities, expressing specific orbital configurations, such as alignment or quadrature. For example, the neighboring pair of planets Jupiter and Saturn induces periodic variations in planetary angular moments of about $\sim$12, $\sim$29, $\sim$18 and $\sim$41 years and the Uranus-Neptune pair induces variations of about $\sim$84, $\sim$165, $\sim$81 and $\sim$250 years. These pairs can be grouped together to obtain another set of periodicities, and so on (see \cite{Morth1979,Lopes2021a,Courtillot2021}). Thus, a 270-year oscillation can be associated with the Uranus-Neptune pair, the Jose cycle (\cf \cite{Jose1965}) with Uranus, the 30-year cycle (unnamed, \cite{Usoskin2017}) with Saturn, the Hale cycle (\cf \cite{Hale1919}) with the Jupiter-Saturn pair, and finally, the famous Schwabe cycle (\cf \cite{Schwabe1844}) with Jupiter. Hence, these Jovian planets and their combinations act through their angular momentum on the solar spots of our star, as illustrated by \blue{Courtillot et al.} (\cite{Courtillot2021}) who used these results to predict the upcoming solar cycle. These commensurabilities result in closely grouped periodicities (\cf \cite{LeMouel2020a}), explaining in particular the variations of the Schwabe cycle between 9 and 14 years since 1700, with an average of 10.8 years. These orbital moments act as an external force on the Sun, and possibly on the Earth itself, and on volcanic activity in particular, as we will examine in this article.

\section{Data and Method \label{sec:04}}
	\subsection{Volcanic eruptions and sunspots \label{sec4.1}}
	
	First, we analyze the datasets of global volcanic eruptions and solar sunspot activity. This link has been studied extensively in the literature, leading to the assumption of a causality relationship between the two phenomena (\eg \cite{Stothers1989,Strestik2003,Herdiwijaya2014, Vasilieva2020}).  Similar studies conducted with other datasets, such as polar movements, have instead concluded common influence (\cf \cite{Courtillot2021,Lopes2021a,Lopes2023}). By jointly analyzing global volcanic eruptions and solar sunspots and proceeding step by step, we aim to better understand this interaction. To do so, we focus on a period of 322 years from January 1700 to February 2022, which begins with the start of the time series of the ISSN.
	 
	The catalog of volcanic eruptions (\cite{venzke2013}) lists 5829 events over the period of interest. It includes a majority (55\%) of moderately explosive events associated with a Volcanic Explosivity Index (VEI) of 2 (see \cite{Newhall1982}), 28\% of effusive eruptions ($0 \leq \textbf{VEI} \leq 1$), 14\% of moderate to cataclysmic eruptions ($3 \leq \textbf{VEI}$) and 3\% of unknown VEI. While these estimates provide general information on volcanic phenomena between 1700 and 2022, we did not consider them to calculate the number of volcanic eruptions per year. In reality, we considered each of these events as an expression of a disruption of the magmatic system balance at a specific location on Earth and at a specific time. These worldwide volcanic eruptions can also be considered as a kind of global parameter highlighting the dynamics of our planet, which is consistent with our global-scale approach. Moreover, the respective analyses of "events with $\textbf{VEI}  \geq 2$" and "all events" conducted in \blue{Dumont et al.} (\cite{Dumont2022}) have led to the detection of similar decadal periodicities.

\begin{figure}[H]
    \begin{center}    
         \includegraphics[width=0.85\columnwidth]{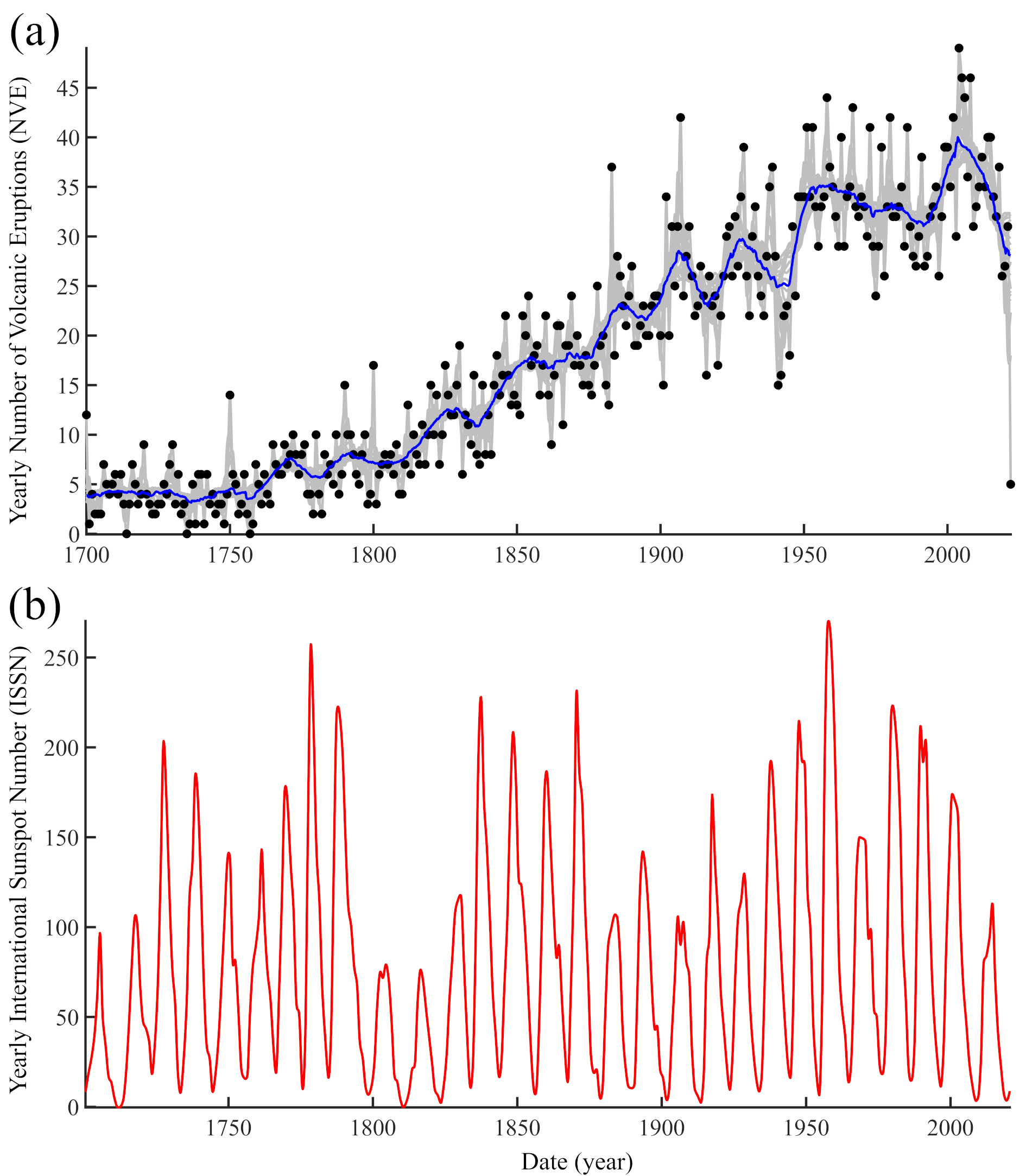}	
         \caption{Number of worldwide volcanic eruptions (NVE) (a) and international sunspot number (ISSN) (b) spanning 1700-2022. (a) The black dots represent the number of worldwide volcanic eruptions per year, connected by the gray curves representing the bootstrapped series, the blue curve representing their median. (b) Temporal evolution of the yearly number of sunspots since 1700. }
         \label{Fig:02}
	
\end{center}
\end{figure}

 	We have taken into account a time interval for which the catalog of volcanic eruptions is only partially complete, which is a critical aspect that must be considered. It is estimated that for $\textbf{VEI} \geq 3$, the GVP catalog is complete since 1960 and for $\textbf{VEI} \geq 4$, since 1820 (\cf. \cite{Gusev2008}). As highlighted by various studies (\eg \cite{Guttorp1991,Simkin1993,Gusev2008,Gusev2014}), a significant increase in the number of volcanic eruptions appears in the second half of the 18th century (Figure \ref{Fig:02}a), resulting in a general upward trend (Figure \ref{Fig:05}a). This trend is mainly related to a more systematic reporting of eruptive events worldwide rather than a real increase in global volcanic activity (\cite{Simkin1993}).  Furthermore, it reverses around the year 2000 (Figures \ref{Fig:02}a and \ref{Fig:05}a), suggesting that 1) over the past two decades, the catalog of global volcanic eruptions is likely to be nearly complete, and 2) this global eruptive activity may not be associated with a stationary process, as generally assumed (\eg \cite{Gusev2008,Marzocchi2012,Gusev2014}. In this study, we will focus on the oscillations composing the time series.
 
	We used the bootstrap technique (\cf \cite{Efron1994}) to construct what we consider as the most representative temporal evolution of volcanic activity over the past 322 years. A thousand iterations were performed to randomly sample the annual number of eruptions listed in the catalog, thus avoiding biases that could be introduced by a time window using a moving average. The various sub-samples range from a minimum of 2 years to a maximum of 40 years and define a family of curves, with the median curve presented in Figure \ref{Fig:02}a. In this study, we considered the median curve as the reference for these temporal variations and will refer to it as NVE.
	
	The ISSN has been measured for more than three centuries to monitor solar activity, with the first yearly records going back to 1700. We used the yearly mean total sunspot number (Figure \ref{Fig:02}b) from the World Data Center SILSO, Royal Observatory of Belgium, Brussels \footnote{http://www.sidc.be/silso/datafiles}.
	
	\subsection{Time-series analysis \label{sec4.2}}	
	The time series of ISSN and NVE were analyzed using iterative Singular Spectrum Analysis (iSSA). For more details on the SSA method, you can refer to \blue{Golyandina and Zhigljavsky} (\cite{Golyandina2013}). For the properties of Hankel and Toeplitz matrices, you can consult \blue{Lemmerling and Van Huffel} (\cite{Lemmerling2001}) and for the Singular Value Decomposition (SVD) algorithm, you can refer to \blue{Golub and Reinsch} (\cite{Golub1971}). The periodicity of each component identified by iSSA is then evaluated using the Fourier transform, with uncertainties given by the full width at half maximum of the peak. In the following text, we prefer to use the term "pseudo-period" to emphasize that these oscillations are not pure oscillatory signals.

\begin{figure}[H]
         \begin{center}
         \includegraphics[width=0.8\columnwidth]{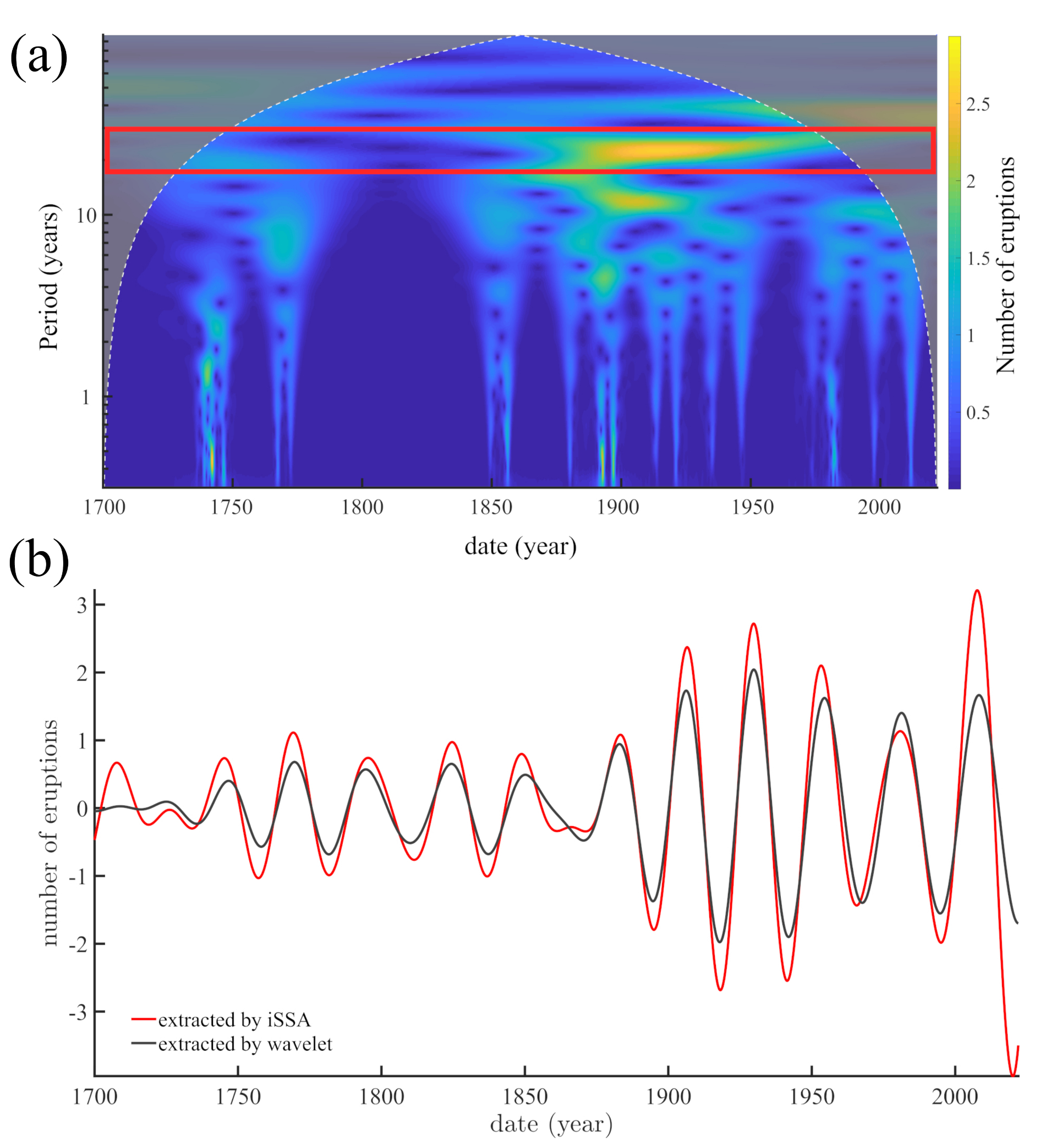}	
         \caption{Extraction of pseudo-oscillations in the NVE time-series: comparison between the continuous wavelet transform and iterative \textbf{SSA} with the example of the 22-year cycle. (a) Scalogram associated with the NVE curve spanning 1700 to 2022. The gray area bordered by the white dashed curve delineates the no-definition sector of the wavelet transform. The red rectangle points out the wavelet coefficients for the 22-year period. (b) The 22-year pseudo-cycle extracted by \textbf{iSSA} (red curve) and the same cycle extracted by continuous wavelet transform (black). The latter corresponds to the wavelet scales boxed in (a).}
         \label{Fig:03}
         \end{center}
\end{figure}

	We also provide a comparison of the 22-year solar cycle extracted by iSSA and by continuous wavelet transform analysis   (\cf \cite{Gibert1998}), a widely used technique in the analysis of solar spot time series  (\eg \cite{Frick1997}).We considered a Morse wavelet which is particularly well-suited for signals with variable spectra (\cf \cite{Olhede2002,Lilly2012}). The scalogram of NVE is shown in Figure \ref{Fig:03}a. We perform the comparison between the two techniques for the 22-year period which represents an important component in the NVE time series (Figure \ref{Fig:03}). Based on the redundancy of information of the wavelet kernels, it is possible to extract the peak corresponding to a specific periodicity within the time scale of interest to reconstruct the ad hoc signal (\cf \cite{Gibert1998}). The difficulty in applying such an approach lies in the spreading of energy over multiple time scales, as illustrated in the 22-year period in Figure \ref{Fig:03}a. This effect can be reduced using a reassignment method (\eg \cite{Auger1995}) to optimize the reconstruction of the period of interest. This would require an additional processing step that we have not implemented in the context of this comparison. In our case, we relied solely on the inverse wavelet transform to reconstruct the pseudo-oscillation corresponding to the peak's maximum energy.

\newpage
The application of iSSA to the NVE time series has allowed us to directly extract various pseudo-cycles, including the one identified with the Scalogram and centered around 22 years. The resulting component is shown in Figure \ref{Fig:03}b and overlaid with the one obtained from continuous wavelet analysis. This figure reveals excellent agreement for the pseudo-oscillations extracted by both techniques, iSSA and continuous wavelet transform, confirming the reliability of the results obtained by iSSA (which is not only faster and easier to implement, but also more precise than wavelet analysis).  

\section{Pseudo-oscillations detected in the \textbf{NVE} and \textbf{ISSN} \label{sec:05}}

\begin{figure}[H]
         \begin{center}
         \includegraphics[width=0.8\columnwidth]{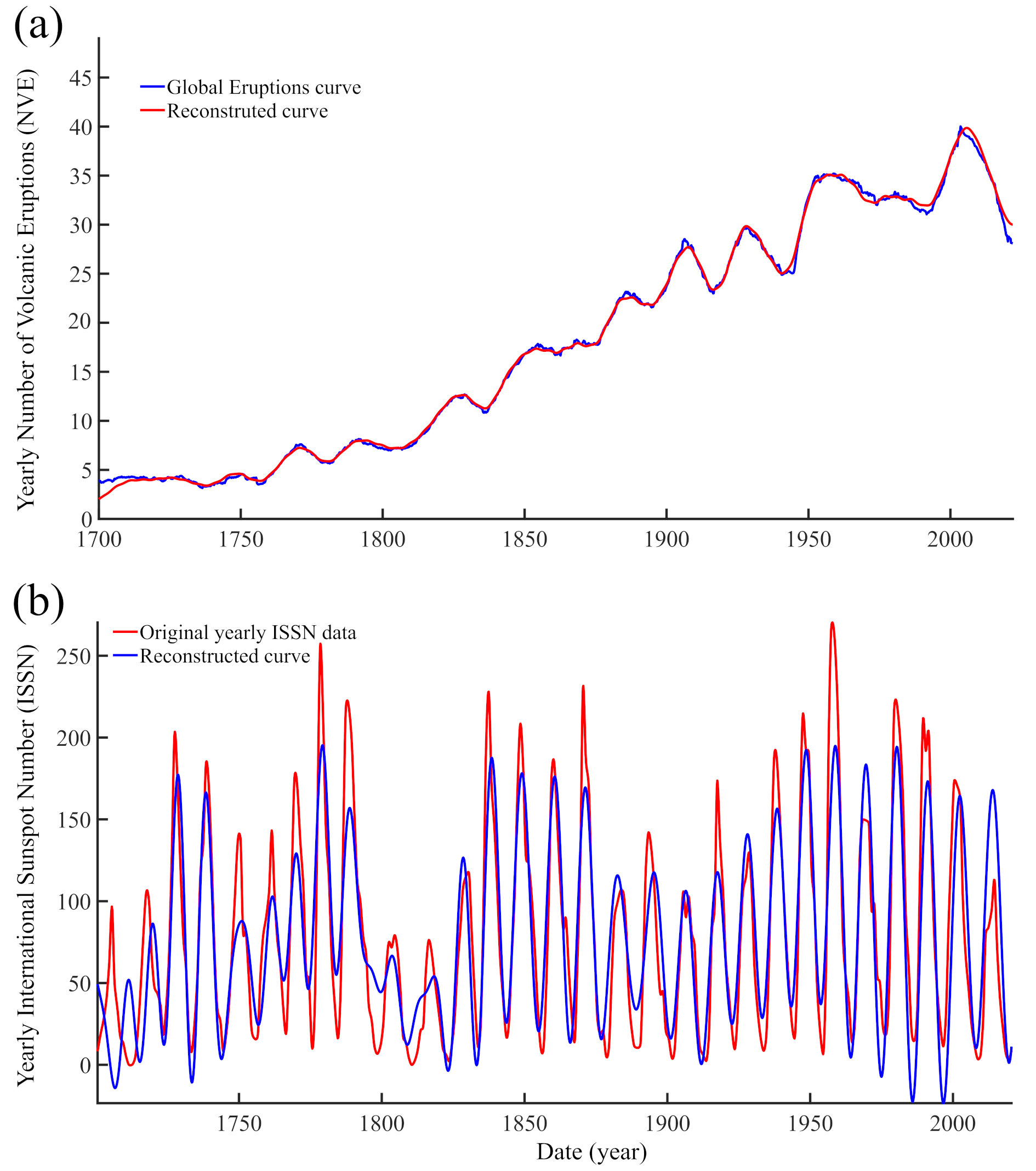}	
         \caption{Original and Reconstructed time-series of NVE (a) and ISSN (b). (a) The NVE curve is represented by the median of bootstrapped series (blue curve aforementioned as original time-series) and its reconstruction (red curve). (b) The original ISSN time-series is shown in red and its reconstruction in blue. See text for more details. }
         \label{Fig:04}
         \end{center}
\end{figure}

\begin{figure}[H]
         \includegraphics[width=1\columnwidth]{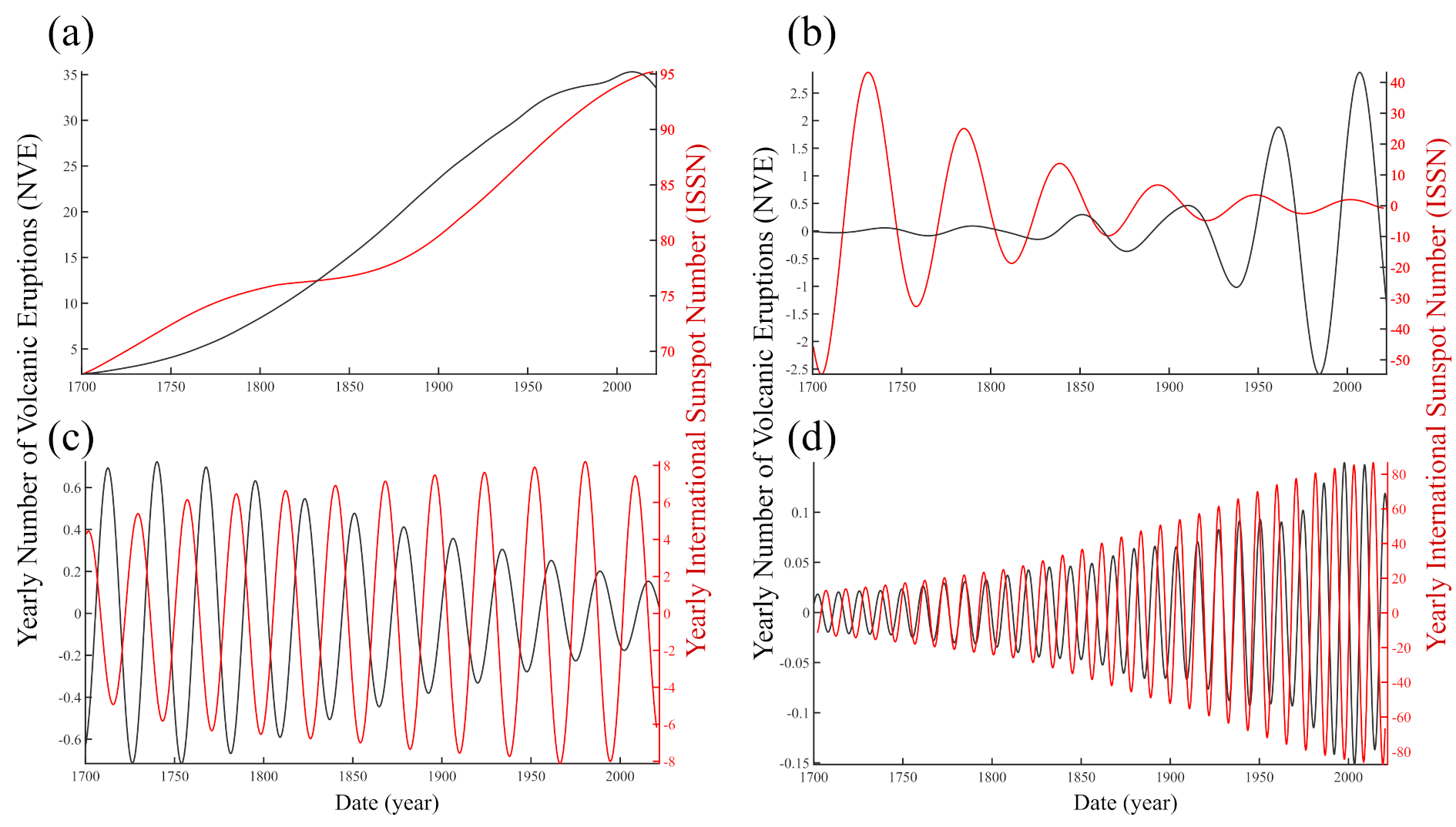}	
         \caption{Trends and  some quasi-periodic components present in both the NVE (black) and ISSN (red) extracted using \textbf{iSSA}. (a) The trends of the NVE and ISSN time-series, (b) the $\sim$50-60-year, (c) the $\sim$27-year and (d) the $\sim$11-year pseudo-periodic components.}
         \label{Fig:05}
\end{figure}

	We have identified thirteen components in the NVE time series using iSSA. These correspond to the first thirty-five components of the original time series. In addition to the trend, twelve periodic components with durations ranging from 4.91 $\pm$ 0.05 to 132.32 $\pm$ 37.57 years have been identified (Table \ref{Tab:01}). They account for a total variance of 77.4\% of the NVE curve, allowing for the reconstruction of most of its variations (Figure \ref{Fig:04}a). Similarly, we analyzed the ISSN time series and found seven pseudo-periods ranging from 8.4 $\pm$ 0.2 to 90.9 $\pm$ 17.5 years, in addition to the trend, using the first fifteen components (Table \ref{Tab:01}). They account for 79.5\% of the total variance (Figure \ref{Fig:04}b), as previously shown in the works of \blue{Le Mouël et al.} (\cite{LeMouel2017,LeMouel2020a}) and \blue{Courtillot et al.} (\cite{Courtillot2021}). Among these seven periodic components, four of them are common to those extracted from NVE. The four common periodicities are all less than 30 years, with periods of approximately $\sim$9, $\sim$11, $\sim$13 and $\sim$27 years (Table \ref{Tab:01} and Figure \ref{Fig:05}).

\begin{table}[H]
         \includegraphics[width=1\columnwidth]{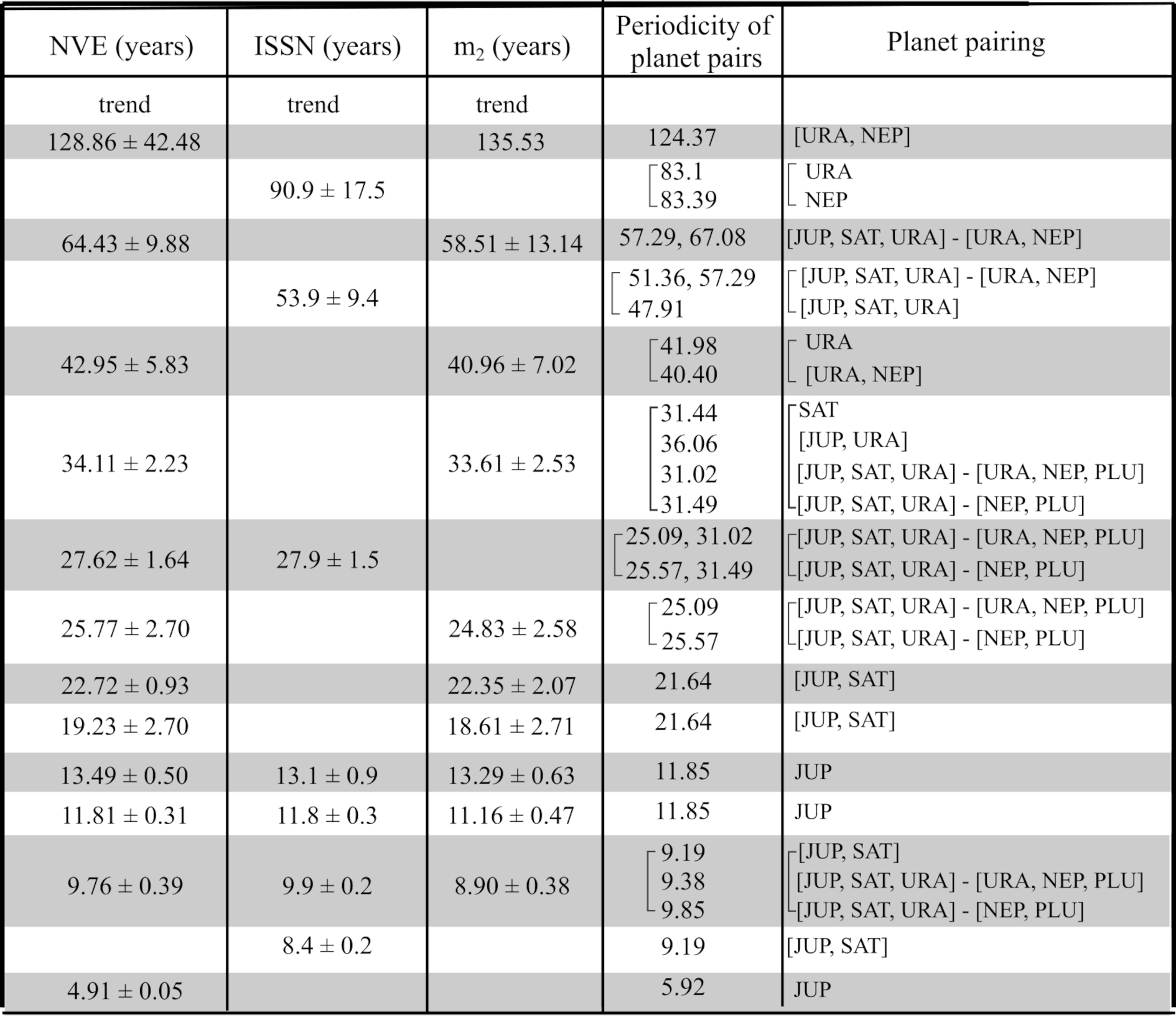}	
         \caption{List of common periodicities extracted from the time-series of NVE, ISSN and $m_2$ component of the PM, using \textbf{iSSA} for the 1700-2022 time interval. All periodicities in the NVE are also present in the $m_2$ time-series, except that of $\sim$27 and $\sim$5 years. The last two columns indicate the combination of planet pairing and the associated periodicities as reported in Courtillot et al., (\cite{Courtillot2021}) and Lopes et al., (\cite{Lopes2021a}). The planets are listed as follows: JUP for Jupiter, NEP for Neptune, PLU for Pluto, SAT for Saturn, URA for Uranus. The brackets separate different pairing.}
         \label{Tab:01}
\end{table}	
	
We now present some of the common pseudo-cycles found in NVE and ISSN time series. First, trends extracted from NVE and ISSN time series show a monotonous increase from 1700 to $\sim$2000 for both datasets, followed by a plateau starting from 2000 (Figure \ref{Fig:05}a). Although their uncertainties are about ten years (Table \ref{Tab:01}), the amplitude modulations of the approximately 60-year and 50-year cycles extracted from NVE and ISSN data are clearly different: while the amplitude of the approximately 50-year component of the solar cycle decreases over time, the amplitude of the approximately 60-year component of NVE increases (Figure \ref{Fig:05}b). This approximately 60-year cycle is well known in climate studies (\eg \cite{Scafetta2010,mazzarella2012,Courtillot2013}).	
	
	In Figure \ref{Fig:05}d, the well-known 11-year solar cycle, also known as the Schwabe cycle (\cf \cite{Schwabe1844}), shows an increase in amplitude in both time series. This similar behavior of amplitudes is accompanied by phases that occasionally synchronize, such as between 1760 and 1800 or between 1910 and 1950. Finally, we illustrate the pseudo-oscillations of 27 years extracted by iSSA from the NVE and ISSN time series (Figure \ref{Fig:05}c). The amplitude variations for both time series do not show as clear correlation as in the cases of approximately 50-60 year cycles and 11-year cycles.

	These common periodicities could explain the numerous correlations obtained between solar and volcanic activity that are discussed in the literature (\cf \cite{Lyons1899,Jensen1902,Schneider1975,Mazzarella1988,Mazzarella1989,Stothers1989,Strestik2003,Herdiwijaya2014, Vasilieva2020}).We have found that several periodicities are common to the NVE and ISSN time series, suggesting a possible link between the two. However, this relationship varies from one cycle to another and evolves over time. This is illustrated with the approximately 50-60 year cycle (Figure \ref{Fig:05}b)  whose amplitude evolves in opposite directions for the components extracted from the NVE and ISSN, or with the 11-year cycle, whose modulation evolves similarly but whose phase apparently synchronizes every $\sim$150 years (Figure \ref{Fig:05}d). Considering that not all the pseudo-cycles detected in the NVE are present in solar activity (Table \ref{Tab:01}) and that the common periodicities have also been detected in other geophysical measurements (\eg \cite{Schlesinger1994,Scafetta2016,Lopes2017,LeMouel2019a,LeMouel2020a}),  we hypothesize that a single common external forcing could explain both the different and common oscillations of these two parameters, NVE and ISSN, whose physics are completely different (\eg \cite{Bank2022,Lopes2023}).

\section{Comparison between \textbf{NVE} and \textbf{PM} \label{sec:06}}
	In \blue{Laplace}’s theory (\cite{Laplace1799}), it is clearly stated that celestial bodies comprising our solar system carry energy that is transferred through exchanges of angular momentum to the Sun and to the Earth's rotation axis. In this section, we examine the periods present in the polar motion (PM) that have already been extracted by \blue{Lopes et al.} (\cite{Lopes2021a}), as well as detected in the length of day (lod) by \blue{Le Mouël et al.} (\cite{LeMouel2019b}). With the exception of the free Chandler oscillation ($\sim$1.19 years) and the annual forced oscillation (1 year), all the periods detected in the variation of day length are also present in the polar motion (Table \ref{Tab:01}). The periods of 1.0 year and ~1.19 years cannot be detected in the variation of day length because annual sampling has been used. This does not question their existence; for example, the annual component has already been suggested in the analysis of regional volcanic activity (\eg \cite{Mason2004}).

\begin{figure}[H]
        \begin{center}
         \includegraphics[width=0.7\columnwidth]{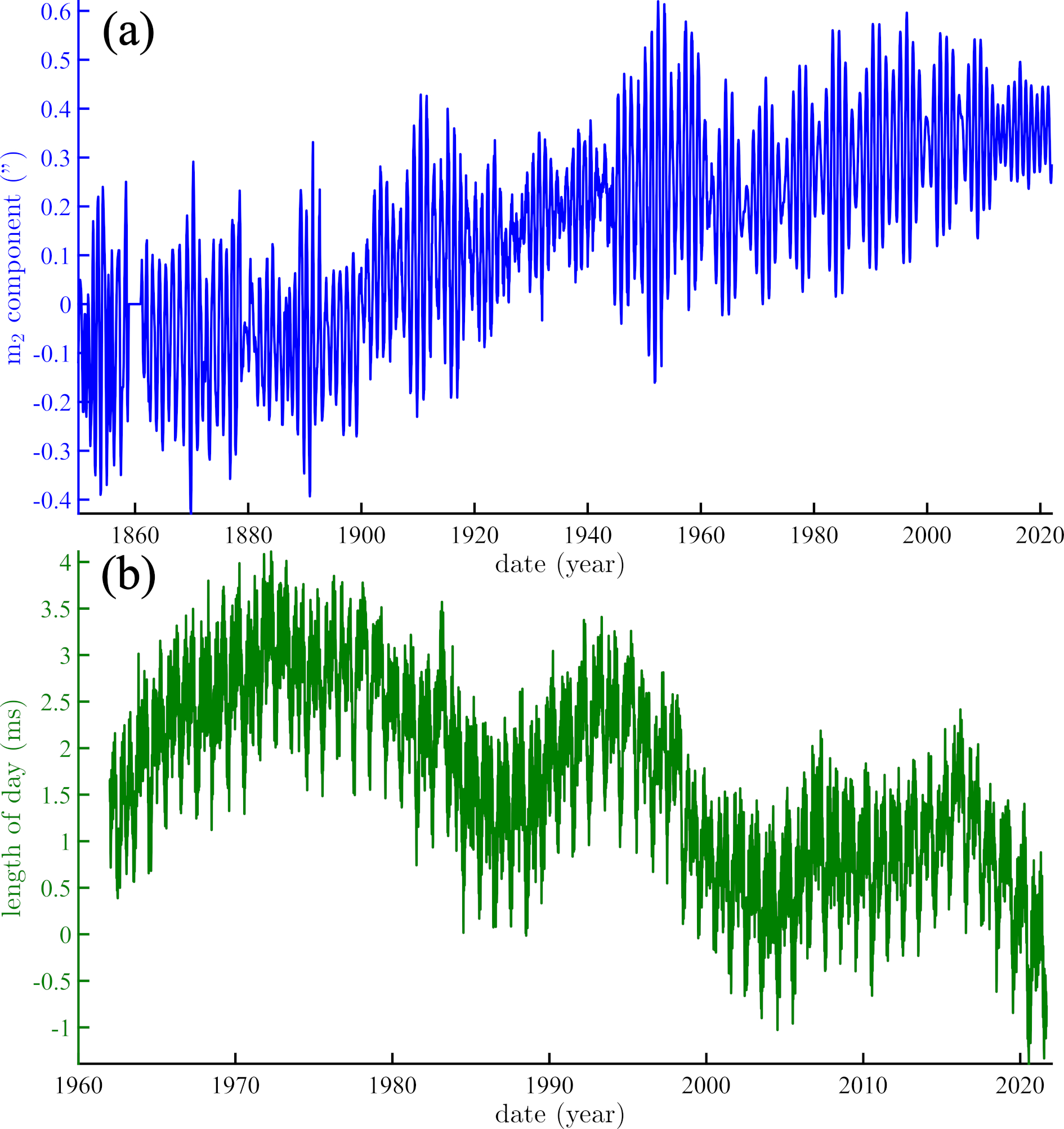}	
         \caption{Temporal evolution of the $m_2$  component of \textbf{PM} since 1846 (a) and the \textit{lod} since 1962 (b) .}
         \label{Fig:06}
        \end{center}
\end{figure}

	Now let's look at the common periodicities extracted from the variation of day length and polar motion, specifically in the E-W component of polar motion called m$_{2}$ (Figure \ref{Fig:06}a). Although this component varies very similarly to component m$_{1}$ (the N-S component), it shows the largest variations of the two polar motion coordinates, which is why we chose to focus on this component. The time series of polar motion starts in 1846, and we present the components extracted by iSSA for the variation of day length and polar motion over the past 176 years, as well as the evolution of their phase lag over time (Figure \ref{Fig:07}). We have also considered day length to complement some observations. Since day length is measured by satellites, this time series is much shorter than the others, starting only in 1962 (Figure \ref{Fig:06}b). Therefore, the comparison between the components of day length variation and day length will be presented over the last sixty years. We draw the reader's attention to the inversion of the axis representing the component in Figures \ref{Fig:07}b-i. Positive values of m$_{2}$ indicate an east-west direction, a sign convention that has no physical significance. The inversion of the axis of the components allows for a better evaluation of the different correlations. In the following, we present the pseudo-periodicities extracted in decreasing order.

	As observed in Figure \ref{Fig:07}a, the two trends extracted from the NVE and PM data show an overall increase over the 176-year interval, although the PM trend begins after a low plateau that ends around 1880. Figure \ref{Fig:07}b represents a cycle centered around $\sim$130-140 years, with a period of 128.86 $\pm$ 42.48 years for NVE and 135.53 years for the $m_2$ component.  It should be noted that there are no uncertainties provided for the PM period (Table \ref{Tab:01}). This is due to the relatively short time interval considered for the time series analysis, which is not sufficient to capture such a long component. Considering the uncertainty in the NVE data, which is approximately 40 years, it can be suggested that this period corresponds to a cycle of approximately 160 years  \cf \cite{Jose1965}). Figure \ref{Fig:07}b  shows that these two components, NVE and $m_2$, are in phase, and although only two oscillations are observed for 1846-2022, their amplitudes evolve similarly, decreasing over time.  	

	A cycle of approximately 60 years has also been extracted, centered at 64.43 $\pm$ 9.88 years for NVE and 58.51 $\pm$ 13.14 years for $m_2$ (Figure \ref{Fig:07}c). As for the cycle of approximately 130-140 years, this cycle for the NVE and PM components is also highly correlated. For this pseudo-cycle and subsequent ones, we calculated the instantaneous phase shift using the phase differences between the Hilbert transforms of NVE and $m_2$. The phase shift for the components of about 60 years varies smoothly, averaging about 1 year. Such a phase shift is small compared to the pseudo-cycle of about 60 years to which it is associated, indicating that this component for NVE and $m_2$ can be considered in phase. In Figure \ref{Fig:07}d, we show a pseudo-cycle of about 40 years extracted in both NVE and $m_2$, approximately  42.95 $\pm$ 5.83 and 40.96 $\pm$ 7.02 years, respectively. The phase correlation appears clearer when it is integrated over time, a transformation for which we will provide further explanations later in this section. We also calculated the continuous phase shift and obtained an average value of about two years between these two components, confirming that NVE and the integrated component of $m_2$ are in phase for the considered time interval. The time series of NVE and $m_2$ include a component of about 25 years whose amplitudes vary in opposite directions (Table \ref{Tab:01} and Figure \ref{Fig:07}e). As for the periodicity of about 40 years, we considered the integral of the component on time to compare it with NVE. Thus, the approximately 25-year component of NVE is shifted by about 3.5 years in 1846, a phase shift that is relatively stable until around 1960 and is in the order of quadrature for the 25-year cycle. This suggests that these two pseudo-cycles are mainly linked by a phase quadrature, which represents a special characteristic that we will explain later in this section.
	
	In Figure \ref{Fig:07}f (left), we present the pseudo-cycle of approximately 19 years with a period estimated at 19.23 $\pm$ 2.70 years for NVE and of 18.61 $\pm$ 2.71 years for $m_2$ component. The amplitude of both components is characterized by a strong increase since 1846, with the increase being more pronounced for the PM component, which has almost zero amplitude until 1900. This cycle is well known and corresponds to the large lunar nutation. As for the approximately 25-year cycle, the phase lag is much larger than the uncertainties on the extracted pseudo-cycle, resulting in evolving phase lags within the considered time interval. Moreover, this lunar periodicity is also detected in the length of day (lod), as shown in \ref{Fig:07}f (right), although the lod time series is much shorter, starting in 1962. By proceeding similarly to what we have done for the previous periodicities, but this time for the lod, we show that the phase lag between NVE and the integral of the lod over time is small and relatively stable over time, averaging about 1 year. This suggests that NVE and the integral of the lod over time are almost in phase. With a similar approach for the approximately 11-year periodicity (Figure \ref{Fig:07}g and Table \ref{Tab:01}), we observe that the amplitudes of NVE and the $m_2$ component vary inversely, and their phase lag increases steadily over time (Figure \ref{Fig:07}g, left). This component was also extracted from the lod, with a period of 11.49 $\pm$ 1.57 years. Thus, NVE and the integral of the lod over time are clearly in phase, as shown by the small variations in the phase lag over the period from 1962 to 2022 (Figure \ref{Fig:07}g, right).

\begin{figure}[H]
	    \centering
	    \begin{subfigure}[b]{\textwidth}
    	     \includegraphics[width=1\columnwidth]{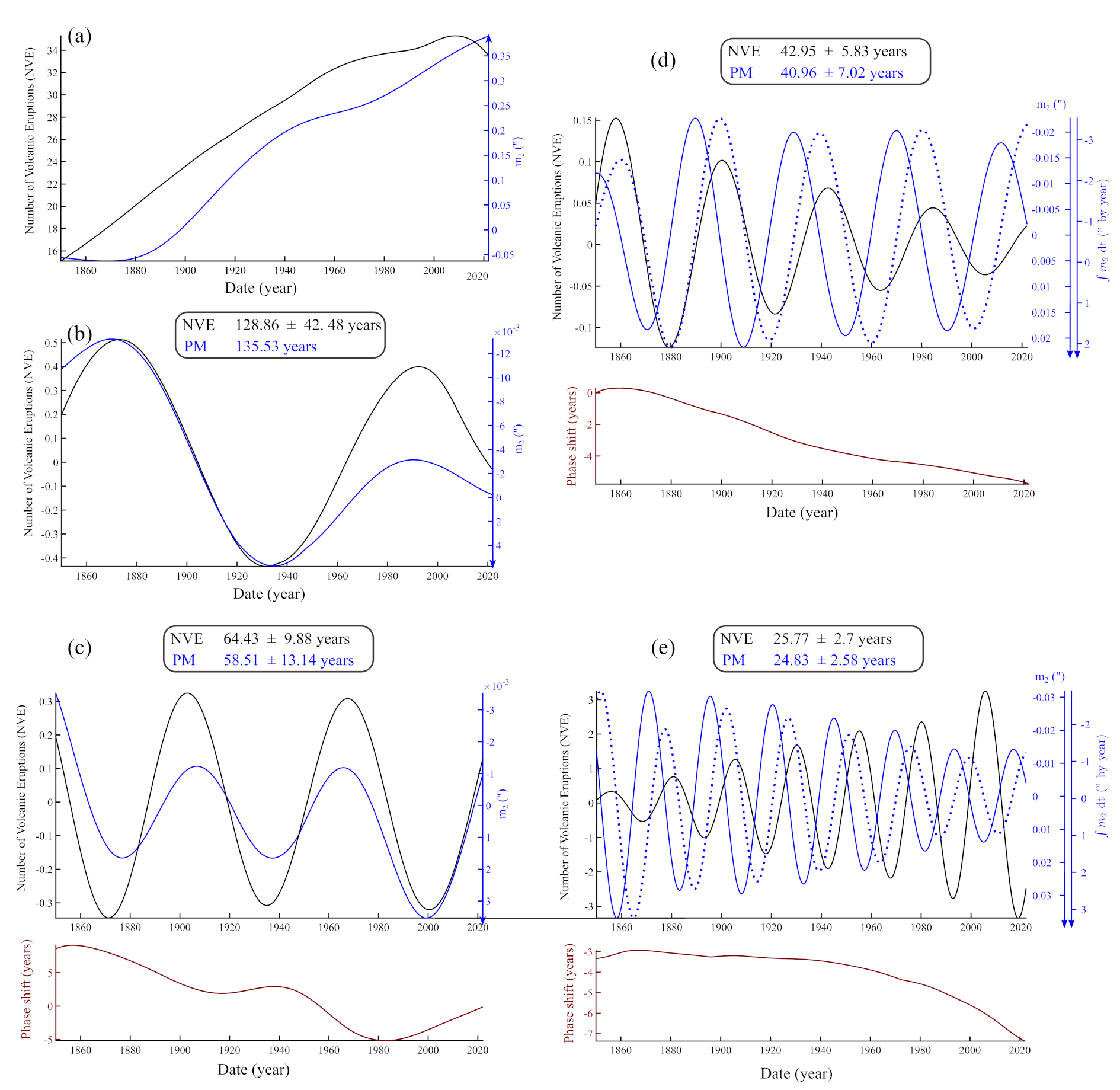}	
        \end{subfigure}   
         \caption{Examples of some common periodicities extracted in the NVE (black curve) and  component of the \textbf{PM} (blue curve, from Lopes et al., \cite{Lopes2021a}) using \textbf{iSSA} with (a) the trends, (b) the $\sim$130-140 year, (c) $\sim$60-year, (d) $\sim$40-year, (e) the $\sim$25-year, (f) the $\sim$19-year, (g) the $\sim$11-year, (h) the $\sim$13-year and (I) the $\sim$9-year pseudo-cycles. For some periodicities (d-i), the integral of $m_2$ over time is also represented by a blue dotted curve. For each periodicity, the phase shift between NVE and $m_2$ component (a-c) or between NVE and the integral of $m_2$ over time (d-i) was calculated and reported in years over the time interval considered. For the $\sim$19 and $\sim$11-year pseudo-cycles, the NVE components was similarly represented but together with that of the \textit{lod} and its integral over time (green full and dotted lines, respectively)}
\end{figure}	
\begin{figure}[H]\ContinuedFloat
	    \centering
		\begin{subfigure}[b]{\textwidth}
			\includegraphics[width=1\columnwidth]{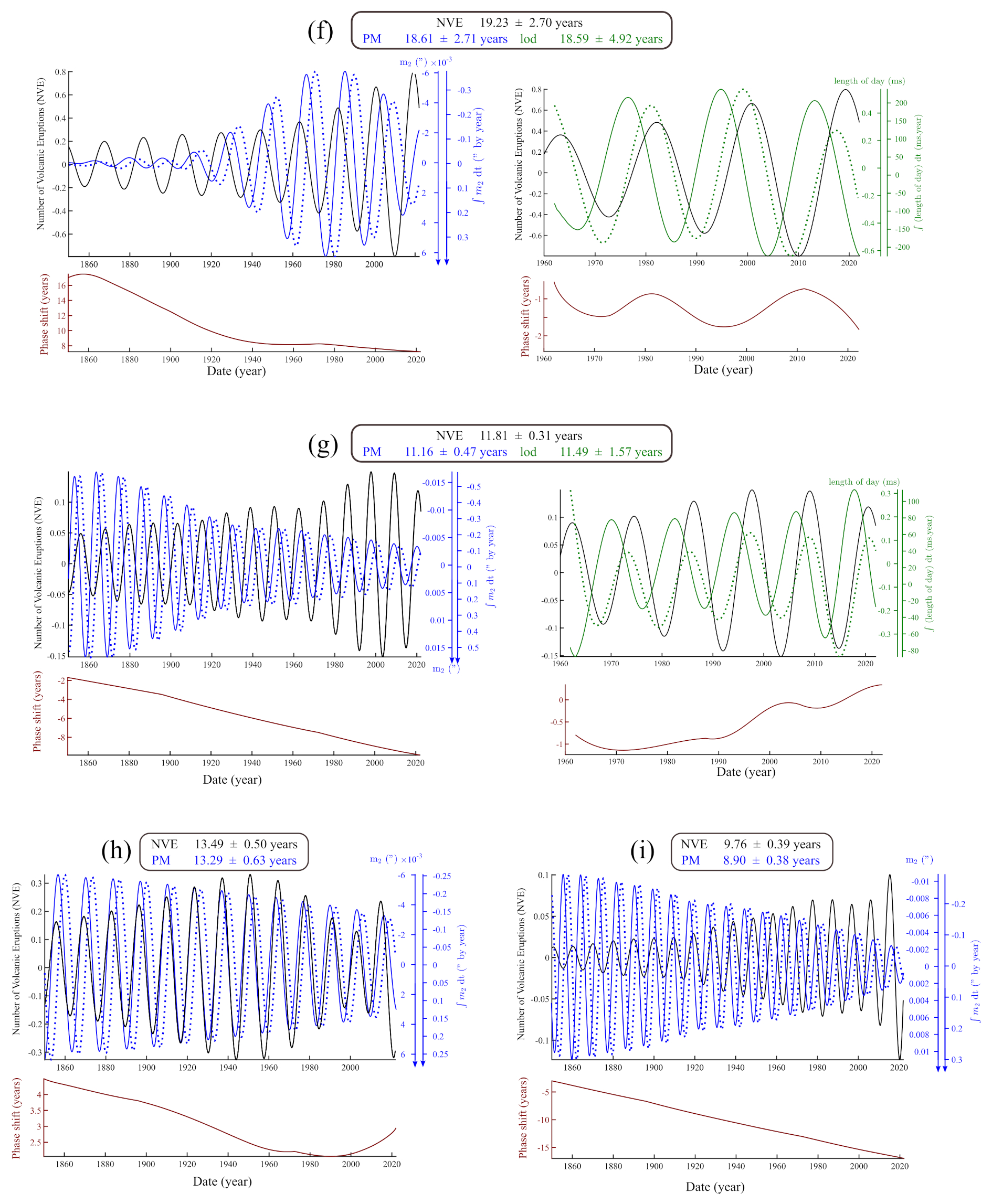}	
        \end{subfigure}           
        
	    \caption{Continued}

        \label{Fig:07}
\end{figure}

	Finally, pseudo-cycles of approximately 13 and 9 years  (Figure \ref{Fig:07}h-i and Table \ref{Tab:01}), are shown for NVE and for the $m_2$ component as well as its time integral. For the pseudo-cycle of approximately 13 years, NVE and $m_2$ are in phase, while NVE and the integral of $m_2$ are in quadrature phase. Their phase lag decreases mainly from 1846 to around 2000, while that of the approximately 9-year periodicity increases. Moreover, this phase lag, being much larger ($>$ 20 times) than the uncertainties of the extracted periodicity, it results in an evolving phase shift between the NVE component and $m_2$.	
	
	In Figures \ref{Fig:07}, we conducted comparative analyses of the extracted periodic components in the NVE, the E-W coordinate of the polar motion (PM) and the length-of-day (lod), which represents the rotational component of PM, for two periodicities. We completed this comparison by including the time integral of PM and lod, whose phase relationships with NVE appear clearer in certain cases. This approach stems from the Louville-Euler equations (system \ref{eq:01}). In reality, PM, lod (related to polar motion variations), and the excitation functions, which are part of mass and stress redistribution, are connected by temporal derivatives. As stated by \blue{Laplace} (\cite{Laplace1799}), demonstrated again by \blue{Lopes et al.} (\cite{Lopes2021a,Lopes2022a}), and explained in more details by a mechanism proposed by \blue{Lopes et al.} (\cite{Lopes2023}), the lod represents the temporal derivative of polar motion. The latter is modulated by exchanges of moments of inertia with other planets in the solar system in a similar way to how the weight of a spinning top interferes with its rotation (\cite{Lagrange1788}). Thus, the stresses resulting from these interactions correspond to the time integral of polar motion, resulting in a quadrature phase relationship, similar to a differentiation link, explaining why we integrated PM and lod over time in Figure \ref{Fig:07}. 

\section{Discussion and Conclusion \label{se:07}}
	The analysis of volcanic activity over the past 300 years is challenging at the scale of individual structures due to the low number of events. However, when considered in the context of a global analysis, volcanic eruptions can provide new insights into worldwide phenomena on Earth that may contribute to triggering eruptions. We have taken a global approach to show that the temporal evolution of the global number of volcanic eruptions is mainly composed of cycles of approximately $\sim$9 to $\sim$130 years. Most of these periodicities are observed in both volcanic activity and the motion of the Earth's rotation pole. Some of these cycles are also found in solar activity (Table \ref{Tab:01}). By using a two-step analysis, we suggest that there is a common external force acting on volcanic eruptions on Earth and solar activity, but not necessarily a causal link between solar spots and volcanic eruptions. Such a link would not be favored, as illustrated by various interpretations suggested in the past (\eg \cite{Lyons1899,Jensen1902,Schneider1975,Mazzarella1988,Mazzarella1989,Stothers1989,Strestik2003,Herdiwijaya2014,Vasilieva2020}). 

	\blue{Dumont et al.} (\cite{Dumont2022}) have identified a link between global volcanic eruptions and the global mean sea level, which is known to exhibit planetary-scale periodicities (\eg \cite{LeMouel2021a,Courtillot2022a}). Such a link can be understood in the case of insular, submarine and coastal magmatic plumbing systems, which are the most abundant (\cite{Dumont2022}): stress variations can be induced by pressure changes imposed by oceanic water mass transport at different timescales. However, for volcanic systems located further inland, the redistribution of oceanic masses likely has a residual effect on their dynamics. Therefore, there must be another causal link that applies to all volcanic systems to explain the common periodicities detected in worldwide volcanic eruptions and solar activity. Recent studies, including \blue{Courtillot et al.} (\cite{Courtillot2021}), \blue{Le Mouël et al.} (\cite{LeMouel2021a}) and \blue{Lopes et al.} (\cite{Lopes2021a,Lopes2022a}), have shown that the temporal evolution of planetary orbital moments is reflected in sunspots and pole motion, as evidenced by their common periodicities. Transfer of energy through planetary orbital moments could represent this external force that is common to volcanic and solar activity.

	In Figures \ref{Fig:07}, we show the components of iSSA that illustrate the common periodicities in NVE and PM (we only present the results for the $m_2$ coordinate of the rotation pole). The most significant evidence of a shared signature and a chain of direct or indirect causality links among the phenomena analyzed in this study comes from Table \ref{Tab:01}. Out of the list of 12 periodicities identified in NVE, only two are apparently absent in $m_2$ ($\sim$27 and $\sim$5 years). All the others fall within the uncertainty range of the extracted components for $m_2$ and NVE. All these shared pseudo-cycles can be associated with a commensurable periodicity that corresponds to the Jovian planets or their combinations in pairs and pairs of pairs (see Table \ref{Tab:01}, \cite{Morth1979,Courtillot2021,Lopes2021a}).
	
	For longer periods than $\sim$19 years, which is associated with the lunar great nutation cycle, good correlations are obtained between the components of the NVE and the polar motion that are either in phase or in anti-phase (Figures \ref{Fig:07}b-e). For shorter periods (Figures \ref{Fig:07}f-i), this phase relationship appears to be lost, and no clear phase relationship between NVE, PM and lod is observed due to higher phase lags. However, the correlations can be improved by considering the time integral of the length of day (lod). Therefore, the phase relationship between NVE and $m_2$ indicates that the period from which the perturbations in the length of day prevail over those of polar motion is that of the great lunar nutation cycle.  

\begin{figure}[H]
         \begin{center}
         \includegraphics[width=0.7\columnwidth]{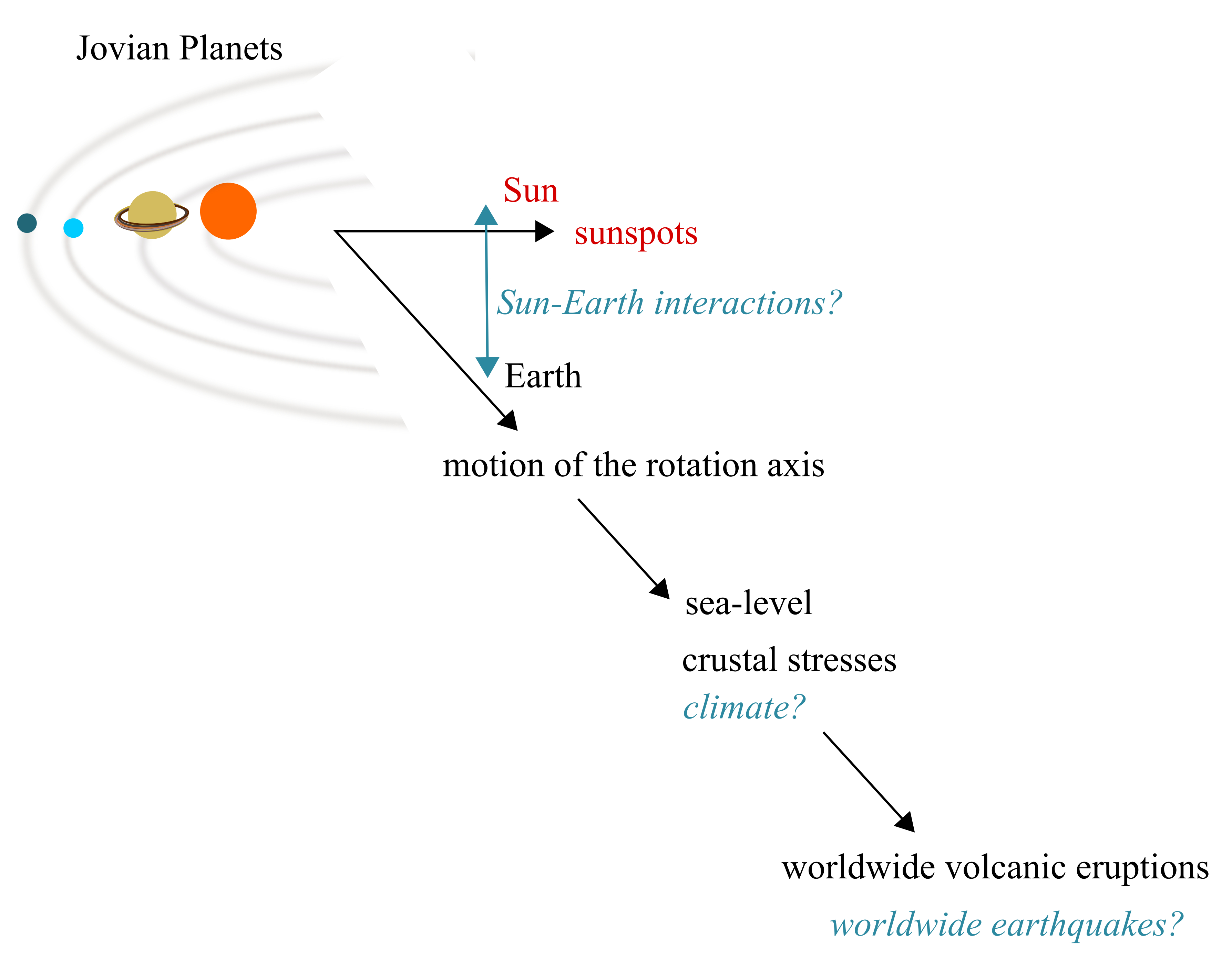}	
         \caption{Cartoon summarizing the action of the Jovian planets on the Sun and Earth dynamics. In blue, a series of other interactions and phenomena that could possibly be affected by the passage of Jovian planets, but not considered in this study.}
         \label{Fig:08}
         \end{center}
\end{figure}

	These observations are consistent with \blue{Laplace}'s theory (\cite{Laplace1799}), particularly the system of linear differential equations of Liouville-Euler (see system \ref{eq:01}). Furthermore, our results suggest that the long periods resulting from the commensurability of planet pairs affect the precession and nutation of the Earth, which describe the major changes in the orientation of the Earth's rotation axis. Shorter periods appear to have a stronger influence on the length of day, i.e., the component that measures the rotation speed around the polar axis. The link between volcanic activity and day length has been explored or suggested in past and recent studies (\cite{Hamilton1973,Stothers1989,Palladino2014,Sottili2015,Tuel2017,Bilham2022}). Thus, two parallel causal chains of forces could be involved, one acting directly on the tilt of the Earth's rotation axis, and the other on its rotation, both modulating crustal stresses and fluid movements that in turn influence the triggering of global-scale volcanic system activity  (Figure \ref{Fig:08}). In light of this study, it is likely that the orbital moments of the Jovian planets and more generally planetary orbits modulate other dynamic processes on Earth, as suggested by other studies (\cite{Zaccagnino2020,Boulila2021,Cionco2021}).  The mathematical and physical mechanisms by which energy transfer to Earth occurs through planetary torques are discussed in  \blue{Lopes et al.} (\cite{Lopes2023}).
	
As suggested for large explosive eruptions in ice core records  (\eg \cite{Tuel2017}), our findings reinforce the idea that volcanic eruptions on Earth cannot be considered as entirely random processes and are partially controlled by external influences related to the ephemerids of celestial bodies in our solar system, acting at multi-annual, multi-decanal, centennial, or even longer timescales. Further investigations would help better understand this interaction as well as the local and regional conditions that cause volcanic systems to respond to these global stress changes, which could potentially contribute to anticipating phases of increased volcanic activity, as suggested by \blue{Bilham et al.} (\cite{Bilham2022}).\\

	In conclusion, we have analyzed the global number of volcanic eruptions (NVE), as well as the number of sunspots (ISSN) and the motion of the Earth's rotation pole (PM) over a period of 322 years (1700-2022). We have demonstrated the presence of a set of common periodicities in global volcanic activity, sunspots, and polar motion. These periodicities range from approximately 5 years to around 130-140 years. We interpret these results within the framework of the Laplace paradigm, the same one used by \Milankovic (\cite{Milankovic1920}) for much longer cycles. With more shared periodicities between NVE and PM than between NVE and ISSN, we propose that these periodic signals are the signature of a common external forcing on solar activity and various global phenomena on Earth, including changes in the tilt of Earth's rotation axis. This results in changes in stress that impact global volcanic activity. The ultimate source of these changes lies in the energy transmitted by the orbital moments of individual Jovian planets and/or their combinations in pairs and pairs of pairs.\\
	\vspace{1cm}

\textbf{Acknowledgement}:
	SD acknowledges Edward Venzke from the Smithsonian Global Volcanism Program for his support in data access. This work was  co-financed  by the FEDER funds through the Programa Operacional Competitividade e Internacionalização - COMPETE 2020 and  through national funds through the  Portuguese Fundação para a Ciência e a Tecnologia (FCT - I.P./MCTES) as part of the SHAZAM (PTDC/CTA-GEO/31475/2017  - POCI-01-0145-FEDER-031475)  project and  by the FCT I.P./MCTES through national funds (PIDDAC) – UIDB/50019/2020- IDL and PTDC/CTA-GEF/6674/2020 - RESTLESS. SD acknowledges the FCT for its financial support through the contrat 2021.00876.CEECIND as well as the FCT and European Union for their ﬁnancial support through the postdoctoral fellowship SFRH/BPD/117714/2016, co-ﬁnanced by the Ministério da Ciência, Tecnologia e Ensino Superior (MCTES), Fundo Social Europeu (FSE), and Programa Operacional Regional Centro (Centro 2020). 			

\newpage	
\bibliographystyle{ieeetr}
\bibliography{volcan}
\end{document}